\documentclass[12pt,article]{amsart2000}
\usepackage{amsfonts}
\RequirePackage{amsmath2000}

\newtheorem{Def-sym}[thm]{Definitions-symbols}

\newcommand{\al} {{\alpha}}

\begin{document}

\pagestyle{myheadings} \markboth{\centerline{\small{\sc Anastasios
Mallios}}}
         {\centerline{\small{\sc $K$-Theory of topological algebras and second quantization}}}

\title{\bf {\boldmath $K$}-Theory of topological algebras and second
quantization}\footnotetext{The present text is an elaborated
version of the talk, delivered by the author at the opening
session of the Intern. Conference on \emph{``Topological Algebras
and Applications''}, Oulu, 2001.}

\author{\bf Anastasios Mallios}

\date{}
\maketitle
\begin{abstract}
Applying the classical \emph{Serre-Swan theorem}, as this is
extended \emph{to topological} (non-normed) \emph{algebras}, one
attains a \emph{classification of elementary particles via} their
\emph{spin-structure}. In this context, our argument is virtually
based on a \emph{``correspondence principle''} of S.A.~Selesnick,
formulated herewith \emph{in a sheaf-theoretic language},
presisely speaking, in terms of \emph{vector sheaves}. This then
leads directly to \emph{second quantization}, as well as, to other
applications of \emph{geometric} (\emph{pre})\emph{quantization
theory}.
\end{abstract}
\setcounter{section} {1} \setcounter{equation} {1}
\bigskip
\hspace{0.5cm} {\bf 0.} Our aim by the present paper is to obtain
a \emph{classification of elementary particles}, that finally
leads to a detour of the so-called first quantization and passage,
instead, directly to \emph{``second quantization''}, which, in
effect, is the main point of \emph{``geometric}
(\emph{pre})\emph{quantization} theory. Consequently, in that
context, \emph{``to find a quantum model of ... an elementary
relativistic particle it is unnecessary ... to quantize} [first]
\emph{the corresponding classical system''}; see
D.J.~Simms-N.M.J.~Woodhouse [26: p. 86]. This is actually in
complete antithesis with what happens, classically. Indeed, in
that case \emph{``... to quantize a field, we have first to
describe it in the language of mechanics''}, see, for instance,
H.~Goldstein [6: p. 370]. On the other hand, \emph{geometric
prequantization} theory is virtually rooted on the standard
\emph{differential geometry of differential} (:
${\mathcal{C}}^{\infty}$-)\emph{manifolds}. Now, the fundamentals
of that classical discipline have been recently formulated within
an entirely \emph{abstract framework}, by employing
\emph{sheaf-theoretic methods}, along with \emph{sheaf cohomology}
theory, yet, \emph{without any use of Calculus}, in the standard
sense of this term, cf. [15], or even [14]. In this connection, an
extremely non-trivial conclusion was the fact that one can
circumvent the underlying space (differential manifold) of the
classical theory and proceed thus, directly, to the study of the
\emph{objects that ``live on the space''}, something that seems to
be of paramount importance for problems in \emph{quantum theory},
that are connected, in particular, with the presence of the
so-called \emph{``singularities''}. Therefore, having this
situation in mind, we aspire, here too, after a
\emph{sheaf-theoretic formulation} of our main results, being thus
in accord with previous considerations, as well as, with their
potential applications, as already described in [15]. Yet,
concerning that latter aspect, pertaining, in particular, to
\emph{gauge theories} of nowadays theoretical physics, we also
refer to A. Mallios [17], where a more detailed account, even of
the present exposition, can still be found (loc. cit. Chapters II,
V).

\bigskip
{\bf 1.} Now, we start with briefly reporting on the manner that;

\bigskip \noindent (1.1) \hfill \begin{minipage}{11cm}
\emph{elementary particles can be classified, according to their
spin structures, in terms of ``vector sheaves''.}
\end{minipage}
\bigskip

First, by the last notion, we formally mean a \emph{sheaf of
modules}, say $\mathcal{E}$, on a \emph{topological space} $X$,
relative to a \emph{sheaf} ${\mathcal{A}}$ \emph{of} (unital
commutative) $\mathbb{C}$-\emph{algebras} on $X$, in such a manner
that $\mathcal{E}$ is, moreover, \emph{locally free of finite
rank}. By the last term, we understand that, every point $x\in X$,
has an (open) neighborhood $U\subseteq X$, on which the
restriction of $\mathcal{E}$ is a \emph{finite power of}
$\mathcal{A}$, the latter sheaf being also similarly restricted on
$U$; thus, by definition, \emph{one has} (up to an ${\mathcal
{A}}|_U$-\emph{isomorphism of} the ${\mathcal
{A}}|_U$-\emph{modules} concerned)
\begin{eqnarray}      
{\mathcal{E}}|_U = {\mathcal{A}}^{n}|_U .
\end{eqnarray}
The number $n\in \mathbb{N}$ in (1.2) \emph{is} kept
\emph{constant}, throughout $X$, and is called the \emph{rank of}
$\mathcal{E}$, with respect to $\mathcal{A}$, thus, by assumption,
\emph{finite} over all $X$. So we write;
\begin{eqnarray}            
rk_{\mathcal{A}}{\mathcal{E}} \equiv rk{\mathcal{E}} =n\in
{\mathbb{N}}.
\end{eqnarray}

The terminology \emph{vector sheaf} for this, otherwise, classical
type of (sheaf) modules, initiated, in effect, within another
context by S.~Lang [8], is mainly due to special
(differential-)geometric applications, that this sort of sheaves
have in an axiomatic treatment of differential geometry, being
undertaken, as already mentioned above, in [15]. Yet, we assume
here that the type of (\emph{local}) \emph{section algebras} of
$\mathcal{A}$ are \emph{unital commutative} (linear associative)
\emph{algebras} over the \emph{complexes}, thus, in short,
$\mathbb{C}$-\emph{algebras}.

So our first objective herewith is to show that:

\bigskip
\noindent (1.4)\hfill \begin{minipage}{11cm} \emph{states of
elementary particles can be associated with} (\emph{local})
\emph{sections of appropriate vector sheaves}, the latter being
provided by the \emph{sheaves of sections of vector bundles} (over
e.g. the \emph{space-time} $X$); again, the latter correspond to
\emph{finitely generated projective modules} over a
\emph{topological algebra}, that, in effect, \emph{cannot} be
Banach, viz. the $\mathbb{C}$-\emph{algebra},

\medskip
(1.4.1) \qquad \qquad \qquad \qquad ${\mathcal{C}}^{\infty}(X),$

\medskip
with $X$, as before, hence, by definition, a \emph{smooth}~(:$\
{\mathcal{C}}^\infty$-) \emph{manifold}.
\end{minipage}
\bigskip

Now, by referring to an \emph{elementary particle}, we mean, by
applying physical parlance of today, an \emph{``ultimate
constituent of the matter''}, something, of course, that virtually
refers to the state of our present-day knowledge, that is to say,
to a certain particular period of time. Anyhow, we suppose in the
sequel that a (\emph{physical}) \emph{particle} will correspond
(\emph{uniquely}) to a \emph{``particle field''}, or simply to a
\emph{``field''}, by thus employing, herewith, \emph{another
name}, in effect, \emph{of the ``particle'' itself}. We have in
that manner the following, by definition, \emph{bijective
correspondence};
\setcounter{equation} {4}
\begin{eqnarray}         
(physical) \ particle \ \ \longleftrightarrow \ \ (particle) \
field.
\end{eqnarray}
The notion in the target of the preceding bijection may be
considered, nowadays, just to quote A.~Einstein, himself, as an
\emph{``independed not further reducible fundamental concept} (see
[5: p. 140]). As a matter of fact, we assume below that
\begin{eqnarray}   
a \ field \ is \ determined \ by \ its \ states.
\end{eqnarray}
Thus, within our axiomatic framework, \emph{states}, as before,
will be just (\emph{local}) \emph{sections} of a suitably defined
\emph{vector sheaf}, that finally will represent the particular
particle (:~field) at issue. In this context, it is worth
remarking here the conceptual coincidence of the previous terms,
as depicted,  by the following (assumed) bijections (the last one
being, in effect, a theorem);
\begin{eqnarray}      
\qquad \qquad field \ \longleftrightarrow \ states \
\longleftrightarrow \ (local) \ sections \ \longleftrightarrow \
vector \ sheaf.
\end{eqnarray}
On the other hand, as we shall see in the sequel, at the final
stage, \emph{``fields''} will be represented by \emph{pairs},
\begin{eqnarray}     
({\mathcal{E}}, D),
\end{eqnarray}
where ${\mathcal{E}}$ is a \emph{vector sheaf} (on $X$, as above),
whose \emph{rank} depends on the particular \emph{spin} of the
particle concerned (see (1.5)), while $D$ stands for an
${\mathcal{A}}$-\emph{connection} (alias, \emph{``covariant
derivative operator''} \emph{on} the vector sheaf (: of the field,
cf.~(1.7)), under consideration. The previous notion has to do
with the corresponding \emph{``field strength''}, or equivalently,
in view of (1.7), with the \emph{curvature} of the
$\mathcal{{A}}$-connection involved, as above. Hence, finally one
has the following (\emph{bijective}) \emph{correspondence}, being
a basic specification of (1.5), viz. we further get;
\begin{eqnarray}      
(particle) field \longleftrightarrow ({\mathcal{E}}, D)
&\longleftrightarrow& field \ strength\\
& &(\equiv curvature, \ R(D)).\nonumber
\end{eqnarray}

\setcounter{section} {2}
\bigskip
{\bf 2.} To continue, we briefly recall some primary facts about
the \emph{spin of elementary particles}, while we also explain how
this concept can be associated with appropriate \emph{modules},
relative to the aforementioned (topological)
$\mathbb{C}$-\emph{algebra} (cf. (1.4.1)),
\setcounter{equation} {0}
\begin{eqnarray}    
{\mathbb{A}} \equiv {\mathcal{C}}^{\infty}(X).
\end{eqnarray}
Thus, in accord with today's point of view, elementary particles
are tabulated in two classes; i) \emph{bosons}, viz. those having
\emph{integer spin}, or even those satisfying \emph{Bose-Einstein
statistics}, and ii) \emph{fermions}, that is, those with
\emph{half-integer spin}, alias obeying \emph{Fermi-Dirac
statistics}.

On the other hand, for (theoretical) convenience, the elementary
particles, we consider throughout, it is supposed to be
\emph{``free''} (or else \emph{``bare''}), while what happens, of
course, in practice, is these particles to be actually detected
(measured), by us, in a \emph{``dressed''} form, namely, in that
form, they acquire, after all the occasional interactions, they
have had, until our own \emph{experiment-measurement}.
Accordingly, by further employing the corresponding classical
(\emph{Hilbert}) \emph{state space} formulation of \emph{quantum
mechanics}, we are led to consider the following relation, as
associated with the actual \emph{state space} of the physical
system at issue. That is, \emph{one has};
\begin{eqnarray}       
{\check H}_{phys} = {\check H}_{bare} \oplus {\check H}_{etc}\;.
\end{eqnarray}
Before we proceed, we explain, in brief, the \emph{notation
applied in} (2.2): Thus, the first component in the second member
of (2.2) stands for the state space, which the physical system
has, when assumed to be \emph{free} (bare), while the second one
at the same part of (2.2) represents the space of those states,
that the system acquires after any interaction of it with other
systems (:~\emph{``perturbation state space''}). Thus, the outcome
is the first member of (2.2), being virtually the \emph{real}
(viz.~\emph{actual}) \emph{state space}, within which all our
measurements are taking place. Of course, operators in physics are
mainly \emph{unbounded}, therefore, \emph{densely defined}
(\emph{``Hellinger-Toeplitz Theorem''}), a fact encoded in (2.2),
by the `` \v~'' notation; in this connection, see also, for
instance, A.~B\"{o}hm [2: p. 19; (3.28)], or even
E.~Prugove\v{c}ki [22: p. 195, Theorem 2.10, yet p.~193 scholia at
the beginning].

Furthermore, as we shall see presently below, the
$\mathbb{C}$-vector spaces, as appeared in the aforesaid relation,
are, in effect, \emph{modules, with respect to the
$\mathbb{C}$-algebra} $\mathbb{A}$, as above, so that, in view of
our previous remarks on the nature of the first member of (2.2)
(thus, we perform within it \emph{measurements}, which are finally
associated with \emph{``coordinates''}), \emph{we} may
\emph{suppose that} the space
\begin{eqnarray}      
{\check H}_{phys}
\end{eqnarray}
is a \emph{free} $\mathbb{A}$-\emph{module}. Consequently, based
on the very definitions (see also the remarks following (2.4)
below), we conclude that;

\bigskip
\noindent (2.4) \hfill \begin{minipage}{11cm}\emph{each one of
the} remaining \emph{two $\mathbb{A}$-modules} in the second
member of (2.2) is a \emph{projective $\mathbb{A}$-module}.
\end{minipage}

\bigskip
In this connection, to support our previous claim in (2.4), and
thus further explain the notation in (2.2), we still remark that
the \emph{``direct sum''} decomposition in the same relation can
be justified, by employing standard arguments, related with
\emph{quantum scattering theory}, pertaining, for the case at
issue, to the so-called \emph{scattering} (alias $\mathcal{S}$-)
\emph{operator}, which here transforms \emph{``in-states''} (viz.
prepared ones) into \emph{``out-states''} (thus, unprepared ones).

\bigskip
\setcounter{section} {3} {\bf 3.} We come next to justify our
previous assertion, concerning, namely,~that;

\bigskip
\noindent (3.1) \hfill \begin{minipage}{11cm} \emph{the
$\mathbb{C}$-vector spaces} appeared \emph{in} (2.2) \emph{are},
in effect, $\mathbb{A}$-\emph{modules}, with respect to the
$\mathbb{C}$-algebra $\mathbb{A}$, as defined by (2.1).
\end{minipage}

\bigskip
The argument is virtually based on a remark of S.A.~Selesnick [25]
in his relevant discussion on the subject, pertaining to the form
that one gets for the \emph{``operator field''}, which corresponds
to the \emph{second quantization of the Dirac field} (viz. the
\emph{relativistic} aspect of \emph{electromagnetism}). So one
has, for instance, the following expression for the field in
question, viz.
\setcounter{equation} {1}
\begin{eqnarray}         
\phi (x,t) = \sum^{k}_{i=1} u_i (x,t)a_i ,
\end{eqnarray}
where the $u_i$'s are \emph{single-particle wave functions}, while
the $a_i$'s stand for the corresponding \emph{``Lader} (i.e.,
\emph{annihilation}) \emph{operators''}; see, for example,
J.D.~Bjorken-S.D.~Drell [1: p. 49; (13.18), and p. 45; (13.3)], or
even A.~B\"{o}hm [2: p. 19]. Therefore, in other words,

\bigskip \noindent (3.3) \hfill \begin{minipage}{11cm}
\emph{the coefficients of the operators involved in second
quantization} (alias, \emph{quantum field theory}) \emph{are not}
just \emph{complex numbers}, as this happens in the first
quantization (usual \emph{quantum mechanics}), \emph{but}
($\mathbb{C}$-valued) \emph{functions}, in point of fact, elements
of the algebra $\mathbb{A}$, as above; accordingly, \emph{the
$\mathbb{C}$-vector spaces involved} in (2.2) \emph{are} virtually
$\mathbb{A}$-\emph{modules}, as claimed.
\end{minipage}

\bigskip
Yet, another justification of the previous argument comes also
from the very definition of the \emph{operator $*$-algebra}, that
corresponds to the system, under discussion (see e.g. A.~B\"{o}hm;
loc. cit., pp. 10,19).

On the other hand, we have already remarked in the previous
Section 2 (cf. (2.4)) that
\setcounter{equation} {3}
\begin{eqnarray}         
{\check{H}}_{bare} \ is \ a \ projective \ {\mathbb{A}}-module.
\end{eqnarray}
As a matter of fact, we proceed now to see that,

\bigskip
\noindent (3.5) \hfill \begin{minipage}{11cm} \emph{the same
$\mathbb{A}$-module}, as above, can also be assumed to be
\emph{finitely generated}.
\end{minipage}

\bigskip
Indeed, the so-called \emph{symmetry group} of a physical system,
that is, the group that \emph{parametrizes the inherent structure}
(\emph{states}) of the system, under consideration, is usually a
\emph{compact} (\emph{matrix}) \emph{Lie group} (see e.g.
R.W.R.~Darling [3: p. 223]). Now, this group is virtually detected
through its \emph{action on} some ($\mathbb{C}$-)\emph{vector
space}, the so-called \emph{representation space} of the group at
issue. Yet, to employ here recent physical parlance (in effect of
\emph{quantum field theory}) by \emph{``curving''} the preceding,
situation one is led to a ($\mathbb{C}$-)\emph{vector bundle} over
a \emph{topological space} $X$ (e.g. the \emph{space-time}, as in
(1.3)), that is associated with the \emph{principal fiber bundle},
whose \emph{structure group} is, by definition, the above
\emph{symmetry group}, being, by assumption (:\emph{``symmetry
axiom''}), the same at the \emph{quantum mechanical level}, as
well (see, for instance, D.J.~Simms-N.M.~Woodhouse [26: p. 150]).
In point of fact, the representation space of the \emph{``symmetry
group''}, as before, should be in our case the $\mathbb{C}$-vector
space appeared in (2.3), the same being, as already explained, a
\emph{free $\mathbb{A}$-module}, with respect to the
$\mathbb{C}$-algebra $\mathbb{A}$, as in (2.1). Now, based on
standard considerations, referring to topological algebra theory,
see e.g. A.~Mallios [11: p. 131, and p. 82, Lemma 1.1, along with
subsequent comments], the same space, as before, may be
considered, as a \emph{complete locally convex}
($\mathbb{C}$-vector) \emph{space}. Therefore, assuming, as
explained before, that our \emph{symmetry group} $G$ is a
\emph{compact Lie group}, and then considering \emph{irreducible
representations} of it in the previous spaces, one concludes that
these are \emph{finite-dimensional} (\emph{``finiteness
theorem''}; see, for instance, A.~Robert [23: p. 69, Corollary
7.9], or even M.A.~Na\u{\i}mark [20: p. 442, Theorem 4]).
Therefore, \emph{by analogy} with the present case, this allows us
to make the following \emph{assumption}:

\bigskip \noindent (3.6) \hfill \begin{minipage}{11cm}
the \emph{space describing} a \emph{field of bare particle
states}, viz.

\medskip
(3.6.1) \qquad \qquad \qquad ${\check H}_{bare}$\;,

\medskip
as in (2.2), is a \emph{finitely generated projective
$\mathbb{A}$-module}. We call it a \emph{quantum state module}.
\end{minipage}

\setcounter{section} {4}
\bigskip
{\bf 4.} Now, in view of our previous statement in (3.6), as well
as, of the standard classification of elementary particles,
according to their \emph{spin-structure} (cf. Section 2), we
conclude that;

\bigskip \noindent (4.1) \hfill \begin{minipage}{11cm}
(\emph{free}) \emph{bosons} correspond to (\emph{finitely
generated}) \emph{projective $\mathbb{A}$-modules of rank 1};
while, (\emph{free}) \emph{fermions} can be described by
\emph{finitely generated projective $\mathbb{A}$-modules of rank
greater than~1}.
\end{minipage}

\bigskip \noindent
We briefly explain right below a plausible argument, on which the
preceding may be rooted: So we first remark that, by considering a
\emph{projective $\mathbb{A}$-module}, say $M$, \emph{of rank 1},
this is \emph{``locally''} identified with \emph{our algebra}
$\mathbb{A}$, the latter being also \emph{``localized''} in a
similar manner; here we refer, of course, to the standard
\emph{localization theory of} (unital commutative
$\mathbb{C}$-)\emph{algebras}, as $\mathbb{A}$ above, this being,
in particular, a \emph{topological} (non-normed) \emph{algebra}.
Thus, one can transcribe the previous \emph{localization argument}
to an analogous one \emph{of the topological algebra $\mathbb{A}$
and the} (finitely generated projective)
$\mathbb{A}$-\emph{module} $M$, with respect to the
\emph{``maximal ideal space''}, alias \emph{``Gel'fand space''},
yet \emph{``spectrum''} of $\mathbb{A}$, denoted herewith by
$\frak{M}(\mathbb{A})$ (see A.~Mallios [11], or even (5.1) below).
Yet, see A.~Mallios [18] for a complete account of the above.

Therefore, elements of $M$ may be (\emph{locally}) considered as
\emph{symmetric functions}, hence, as appropriate to represent
(still locally) \emph{wave functions, states of bare bosons}, viz.
of \emph{free particles}, obeying \emph{Bose-Einstein statistics}.

On the other hand, \emph{bare fermions}, namely, free particles,
obeying \emph{Fermi-Dirac statistics}, correspond to states, that
can be expressed through \emph{antisymetric wave functions};
consequently, by analogy with the previous
\emph{transcription-representation of the elements of $a$}
(finitely generated projective) $\mathbb{A}$-\emph{module} $M$,
\emph{as ``local functions''}, one has to consider here
(\emph{finite}) \emph{exterior powers of $M$} the latter
\emph{being}, of course, $\mathbb{A}$-\emph{modules of the same
type, as} $M$ itself, that is, \emph{finitely generated projective
$\mathbb{A}$-modules of rank, at least 2}. (In this connection, we
still refer to A.~Mallios [17] for a more detailed account of the
preceding, as well as, to S.A.~Selesnick [25]. Yet, cf. A.~Mallios
[10: Section 1, p. 454 ff]).

\setcounter{section} {5}
\bigskip
{\bf 5.} Our final aim in this and the following section is to
relate our previous conclusion in (3.6) with \emph{vector bundles}
and their associated (\emph{vector}) \emph{sheaves} on $X$, the
latter space being, as hinted at in the preceding, the
\emph{spectrum} of the \emph{topological algebra} $\mathbb{A}$.
That is, one has,
\setcounter{equation} {0}
\begin{eqnarray}    
{\frak{M}}({\mathbb{A}}) \equiv {\frak{M}}
({\mathcal{C}}^{\infty}(X)) =X,
\end{eqnarray}
the last relation denoting, in effect, a \emph{homeomorphism} of
the topological spaces concerned (see A.~Mallios [11: p. 227,
Theorem 2.1]). Now, this interrelation between
$\mathbb{A}$-modules of the previous type (cf. e.g. (3.6)) and
vector bundles on $X$, as in (4.1), is actually based on the
so-called \emph{Serre-Swan theorem} (see e.g. M.~Karoubi [7]). We
can call, therefore, the interrelation at issue, the
\emph{Serre-Swan correspondence}.

In point of fact, the aforementioned \emph{Serre-Swan theorem}
refers, in its standard form, to \emph{finitely-generated
projective ${\mathcal{C}}(X)$-modules}, with $X$ \emph{compact},
corresponding, \emph{bijectively} (viz. the respective
\emph{categories are equivalent}), to (\emph{continuous})
\emph{complex $n$-plane bundles} on $X$, with $n\in \mathbb{N}$,
the \emph{rank of the ${\mathcal{C}}(X)$-modules} concerned. There
is also a version of the same result for (finite-dimensional)
\emph{smooth} (viz. ${\mathcal{C}}^{\infty}$-) \emph{bundles} on
the \emph{compact} $({\mathcal{C}}^{\infty}$-) \emph{manifold} $X$
(see e.g. K.~L{\o}nsted [9: p. 201]). On the other hand, as
already said, the $\mathbb{C}$-algebra $\mathbb{A}$, as in (2.1),
is a \emph{topological algebra}, which \emph{cannot be normable}.
Thus, the above theorem, as generalized to (non-normed)
topological algebras, has the following form, in terms of
\emph{Grothendieck $K$-groups} (we refer to A.~Mallios [10] for
the terminology applied herewith);
\begin{eqnarray}     
K(X)=K({\mathbb{A}}) = K({\mathcal{P}} ({\mathbb{A}})),
\end{eqnarray}
within an \emph{isomorphism of abelian groups}, in such a manner
that,
\begin{eqnarray}     
X \sim {\frak{M}}({\mathbb{A}}),
\end{eqnarray}
viz. $X$ is a \emph{topological space homotopic to}
${\frak{M}}({\mathbb{A}})$, the \emph{spectrum of} $\mathbb{A}$.
In this connection, we still note that;

\bigskip \noindent
(5.4) \hfill
\begin{minipage}{11cm}
the $\mathbb{C}$-algebra $\mathbb{A}$, as in (5.2), is now a
\emph{unital commutative locally $m$-convex $Q$-algebra} (alias, a
\emph{Waelbroeck algebra}).
\end{minipage}

\bigskip
\noindent However, see also (5.13) in the sequel.

On the other hand, we remark that the algebra ${\mathbb{A}}\equiv
{\mathcal{C}}^{\infty}(X)$, with $X$ a \emph{compact}
(\emph{Hausdorff}) \emph{smooth manifold} is a \emph{Waelbroeck
algebra}, as in (5.4), therefore, one thus gets at the
${\mathcal{C}}^\infty$-\emph{analogue of Swan's theorem}, as
above; yet, we still note here that the previous property of
$\mathbb{A}$, viz. of being a $Q$-\emph{algebra, characterizes the
compactness of} $X$ (cf. A.~Mallios [15: p. 371; (11.39)]).
Furthermore, by employing standard terminology, we also refer to
(5.2), by just saying that;

\bigskip \noindent
(5.5) \hfill
\begin{minipage}{11cm}
\emph{any continuous $n$-dimensional $\mathbb{C}$-vector bundle is
algebraic, relative to $\mathbb{A}$, as in} (5.4) (but, see also
(5.13) below).
\end{minipage}

\bigskip \noindent
Now, we can further express (5.2), in terms of the so-called
\emph{``projection operators''}, or just \emph{``projectors''},
namely, \emph{``idempotent''} elements of some algebra of
operators (:\emph{linear endomorphisms}). So if we take an element
\setcounter{equation} {5}
\begin{eqnarray}      
[M] \in K({\mathbb{A}}),
\end{eqnarray}
with $M$ a \emph{finitely generated projective
$\mathbb{A}$-module}, then \emph{one obtains};
\begin{eqnarray}      
M=\ker (\al ),
\end{eqnarray}
\emph{such that}
\begin{eqnarray}        
\al \in M_n ({\mathbb{A}}),\ \ with \ \ \al^2 =\al  .
\end{eqnarray}
Thus, in turn, one defines a \emph{morphism}
\begin{eqnarray}        
{\hat{\al}}: X \times {\mathbb{A}}^{n} \to X \times
{\mathbb{A}}^n,
\end{eqnarray}
\emph{such that}, more precisely, one has
\begin{eqnarray}        
[M] = [\ker (\hat{\al})] \in K(X).
\end{eqnarray}
Furthermore, by an obvious \emph{abuse of notation}, we simply
write,
\begin{eqnarray}       
\xi \equiv (E, \pi ,X)=\ker (\al ),
\end{eqnarray}
where $\xi$ stands for a \emph{continuous finite-dimensional
$\mathbb{C}$-vector bundle} over $X$. We recall here that $X$ is a
\emph{compact} (\emph{Hausdorff}) topological \emph{space}, that
is further assumed to be \emph{homotopic to the spectrum} of the
\emph{topological algebra} $\mathbb{A}$ (see (5.3), (5.4)). In
this connection, we still refer to A.~Mallios [10], for further
details. On the other hand, one can also get at the following
\emph{generalization} of the preceding. Thus, \emph{one obtains};

\bigskip \noindent
(5.12) \hfill
\begin{minipage}{11cm}
(5.12.1) \qquad \qquad \qquad $K(X) = K({\mathbb{A}})$,

\medskip \emph{for any}

\medskip (5.12.2) \qquad \qquad \qquad \quad ${\mathbb{A}} =
\varinjlim {\mathbb{A}}_{\al}$,

\medskip viz. a \emph{topological algebra, inductive limit of
Waelbroeck algebras} (cf. (5.4)), in such a manner that, \emph{one
has};

\medskip (5.12.3) \qquad \qquad \quad $X= \varprojlim X_{\al}
(\equiv {\frak{M}}({\mathbb{A}}_{\al}))$,

\medskip\emph{such that}

\medskip(5.12.4) \qquad \qquad \quad  $K(X_{\al})=
K({\mathbb{A}}_{\al}), \ \ \al \in I$.
\end{minipage}

\bigskip
\setcounter{equation} {12}

\noindent Indeed, \emph{one gets};
\begin{eqnarray}        
K({\mathbb{A}}) &=& K(\varinjlim_{\al} {\mathbb{A}}_{\al})
=\varinjlim_{\al} K({\mathbb{A}}_{\al}) =\varinjlim_{\al}
K(X_{\al})\\ &=& K(\varprojlim_{\al} X_{\al})=K(X).\nonumber
\end{eqnarray}
In this regard, see also e.g. J.~Rosenberg [24: p. 9, Theorem
1.2.5], along with R.G.~Swan [27: p. 214, Theorem 7.1]; see also
A.~Mallios [11] for further technical details.

\setcounter{section} {6}
\bigskip
{\bf 6.} Now, denoting by
\setcounter{equation} {0}
\begin{eqnarray}    
P_n (X)
\end{eqnarray}
the set \emph{of isomorphism classes of smooth $n$-dimentional
$\mathbb{C}$-vector bundles} over the \emph{compact manifold} $X$,
and by
\begin{eqnarray}     
\Phi^{n}_{\mathcal{A}} (X)
\end{eqnarray}
the \emph{set of isomorphism classes of vector sheaves on $X$, of
rank} $n\in \mathbb{N}$, with
\begin{eqnarray}       
{\mathcal{A}}\equiv {\mathcal{C}}^{\infty}_{X},
\end{eqnarray}
that is, the \emph{sheaf of germs of $\mathbb{C}$-valued smooth}
(viz. ${\mathcal{C}}^{\infty}$-) \emph{functions on $X$}, one gets
\begin{eqnarray}       
P_{n}(X)=\Phi^{n}_{\mathcal{A}}(X),
\end{eqnarray}
within a \emph{bijection} (see A.~Mallios [14: Chapt. XI; p. 344,
Theorem 8.2], or even A.~Mallios [12: p. 409, Scholium 1.1]).
Therefore, based further on (5.2), with $\mathbb{A}$ given by
(2.1), and on our previous conclusion in (4.1), \emph{we} finally
\emph{conclude that}:

\bigskip \noindent
(6.5) \hfill
\begin{minipage}{11cm}
\emph{states of} bare \emph{bosons} are represented by (continuous
local) \emph{sections of line sheaves} on $X$ (the latter space
being, for instance, the space-time manifold), while \emph{states
of} bare \emph{fermions}, by similar \emph{sections of vector
sheaves} on $X$, \emph{of rank}, at least, \emph{2}.
\end{minipage}

\bigskip \noindent
As a consequence, the corresponding (\emph{quantum}) \emph{fields}
(see (1.7), (1.9)) can be represented by pairs
\setcounter{equation} {5}
\setcounter{equation} {5}
\begin{eqnarray}      
({\mathcal{L}},D)
\end{eqnarray}
for \emph{bosons}, called \emph{Maxwell fields} (generalizing thus
the case of \emph{electromagnetic field}, cf.~also, for instance,
Yu.I.~Manin [19: p. 71]), while in the case of \emph{fermions}, by
pairs of the form,
\begin{eqnarray}      
({\mathcal{E}},D),
\end{eqnarray}
called \emph{Yang-Mills fields}; in this regard, $D$ stands here
for an ${\mathcal{A}}$-\emph{connection}, in the sense of
\emph{abstract differential geometry} [15], defined on the vector
sheaves, in general, $\mathcal{L}$ and $\mathcal{E}$, as in (6.5),
respectively.

\setcounter{section} {7}
\bigskip
{\bf 7. Concluding remarks.-}  The preceding provides, in effect,
an \emph{axiomatic treatment of elementary particles}, through the
aforesaid pairs, as in (6.6) and (6.7), that is virtually rooted
on \emph{Selesnick's correspondence principle}, as exhibited in
the previous discussion. The above point of view lies also at the
basis of our treatment of \emph{gauge theories}, in terms of
\emph{abstract differential geometry}, as expounded in A.~Mallios
[17]. Yet, within the same vein of ideas, \emph{second
quantization} may be construed, as an attempt to look at
\emph{Schr\"{o}dinger's} (\emph{wave}) \emph{equation}, as the
\emph{source}, e\;o\;\;i\;p\;s\;o, \emph{of an} (elementary
particle) \emph{field} (hence, of the
$\mathcal{A}$-\emph{connection} $D$, which is involved, see, for
instance (1.9)) that is, of \emph{the field itself}, and not
merely, as an equation of the vector states in the carrier space
of a particular representation of CCR (:\emph{first
quantization}). On the other hand, a \emph{similar echo},
regarding the meaning of the above form of (elementary)
\emph{particles} $\leftrightarrow$ \emph{fields}, as in (6.6) and
(6.7), can be recognized already in relevant passages of the work
of V.I.~Denisov-A.A.~Logunov [14], as well as, in that of
T.H.~Parker [21]. Yet, as another conclusion of the above
discussion, one obtains that:

\bigskip \noindent
(7.1) \hfill
\begin{minipage}{11cm}
\emph{every} (bare) \emph{elementary particle is
(pre)quantizable}.
\end{minipage}

\bigskip
Further details on this aspect are presented in A.~Mallios [15:
Chapt.~X; p. 293, (5.13), or even [17]. On the other hand, a first
announcement of (7.1) can already be found in A.~Mallios [13: p.
199; (9.3)].

Finally, something that is worth mention here is the aspect that,
following the point of view of the previous discussion, one
appropriately transfers properties of the (underlying) space (e.g.
compactness) to the objects that live on it (e.g.~\emph{``vector
bundles of finite type''}). This point of view has been
systematically advocated in A.~Mallios [15], concerning
fundamental notions and results of the standard differential
geometry on smooth manifolds, while a similar situation can still
be recognized already in the work of L.N.~Vaserstein [28], where a
\emph{generalization of} the classical \emph{Serre-Swan theorem},
as cited in the preceding, is obtained for an \emph{arbitrary
topological space}, by considering, however, a suitable type of
(continuous) vector bundles (viz. such of \emph{``finite type''}).

Thus, in other words, our main motto herewith is that;

\bigskip \noindent
(7.2) \hfill
\begin{minipage}{11cm}
\emph{properties} that were \emph{being considered, thus far}, as
\emph{inherent of the underlying space} (so that the objects, at
issue, that ``live'' on it have the corresponding desired ones),
\emph{are now transferred to the objects themselves}, after, of
course, we have appropriately  \emph{transcribed} these properties
\emph{in an algebraic} (viz. operational-theoretic and, precisely
speaking, \emph{sheaf-theoretic}) \emph{form}.
\end{minipage}

\bigskip
Yet, at the very end, such properties are virtually, encoded in
our \emph{``arithmetics''}, alias \emph{``sheaf of
coeffiecients''} $\mathcal{A}$, that, in turn, can be transported
to the $\mathcal{A}$-\emph{modules} involved. (See, for instance,
the case of an $\mathcal{A}$-\emph{metric}; A.~Mallios [15: Chapt.
IV], [16], [17]).

\end{document}